\documentclass{iau} 
\usepackage{graphicx}

\def\etal{et al.}

\def\teff{\ifmmode T_{\rm eff} \else $T_{\mathrm{eff}}$\fi}

\def\mh{$\mathrm{[M/H]}$}
\def\ltsima{$\buildrel<\over\sim$}
\def\lsim{\lower.5ex\hbox{\ltsima}}

\newcommand{\hii}{H~{\sc ii}}
\newcommand{\ha}{\ifmmode {\rm H}\alpha \else H$\alpha$\fi}
\newcommand{\hb}{\ifmmode {\rm H}\beta \else H$\beta$\fi}
\newcommand{\lya}{\ifmmode {\rm Ly}\alpha \else Ly$\alpha$\fi}

\newcommand{\hep}{He$^+$}

\newcommand{\heii}{He~{\sc ii}}
\newcommand{\Heiiuv}{He~{\sc ii} $\lambda$1640}
\newcommand{\Heii}{He~{\sc ii} $\lambda$4686}

\newcommand{\ga}{\raisebox{-0.5ex}{$\,\stackrel{>}{\scriptstyle\sim}\,$}}

\newcommand{\ebv}{\ifmmode E_{\rm B-V} \else $E_{\rm B-V}$\fi}
\newcommand{\av}{\ifmmode A_{\rm V} \else $A_{\rm V}$\fi}

\def\ergs{erg s$^{-1}$}

\def\msun{\ifmmode M_{\odot} \else M$_{\odot}$\fi}
\def\msunyr{\ifmmode M_{\odot} {\rm yr}^{-1} \else M$_{\odot}$ yr$^{-1}$\fi}
\def\zsun{\ifmmode Z_{\odot} \else Z$_{\odot}$\fi}

\def\lsun{\ifmmode L_{\odot} \else L$_{\odot}$\fi}

\def\mup{\ifmmode M_{\rm up} \else M$_{\rm up}$\fi}
\def\mlow{\ifmmode M_{\rm low} \else M$_{\rm low}$\fi}

\def\aap{A\&A}
\def\araa{ARAA}

\def\apj{ApJ}
\def\apjl{ApJL}
\def\apjs{ApJS}
\def\mnras{MNRAS}

\newcommand{\oh}{\ifmmode 12 + \log({\rm O/H}) \else$12 + \log({\rm O/H})$\fi}
\def\flyf{\ifmmode f_{\rm Lyf} \else $f_{\rm Lyf}$\fi}
\def\pz{\ifmmode P(z) \else $P(z)$\fi}
\def\ki2{\ifmmode \chi^2 \else $\chi^2$\fi}
\def\zphot{\ifmmode z_{\rm phot} \else $z_{\rm phot}$\fi}

\newcommand{\xphot}{\ifmmode x_\gamma \else $v_\gamma$\fi}
\newcommand{\xobs}{\ifmmode x_{\rm obs} \else $x_{\rm obs}$\fi}
\newcommand{\xcmf}{\ifmmode x_{\rm CMF} \else $x_{\rm CMF}$\fi}
\newcommand{\vexp}{\ifmmode V_{\rm exp} \else $V_{\rm exp}$\fi}
\newcommand{\vmax}{\ifmmode V_{\rm max} \else $V_{\rm max}$\fi}
\newcommand{\nh}{\ifmmode N_{\rm HI} \else $N_{\rm HI}$\fi}
\newcommand{\dv}{\ifmmode \Delta v({\rm em-abs}) \else $\Delta v({\rm em}-{\rm abs})$\fi}

\def\fesc{\ifmmode f_{\rm esc} \else $f_{\rm esc}$\fi}

\def\frellya{\ifmmode f^{\rm rel}_{\rm{Ly}\alpha} \else $f^{\rm rel}_{\rm{Ly}\alpha}$\fi}

\def\hii{H{\sc ii}}

\newcommand{\mstar}{\ifmmode M_\star \else $M_\star$\fi}
\newcommand{\muv}{\ifmmode M_{1500} \else $M_{1500}$\fi}
\newcommand{\auv}{\ifmmode A_{\rm UV} \else $A_{\rm UV}$\fi}
\newcommand{\luv}{\ifmmode L_{\rm UV} \else $L_{\rm UV}$\fi}
\newcommand{\lir}{\ifmmode L_{\rm IR} \else $L_{\rm IR}$\fi}
\newcommand{\lbol}{\ifmmode L_{\rm bol} \else $L_{\rm bol}$\fi}
\newcommand{\liruv}{\ifmmode L_{\rm IR+UV} \else $L_{\rm IR+UV}$\fi}
\newcommand{\liroveruv}{\ifmmode L_{\rm IR}/L_{\rm UV} \else $L_{\rm IR}/L_{\rm UV}$\fi}
\newcommand{\nlyc}{\ifmmode N_{\rm Lyc} \else $N_{\rm Lyc} $\fi}
\newcommand{\rholyc}{\ifmmode \rho_{\rm Lyc} \else $\rho_{\rm Lyc} $\fi}
\newcommand{\chion}{\ifmmode \xi_{\rm ion} \else $\xi_{\rm ion}$\fi}
\newcommand{\chioncorr}{\ifmmode \xi_{\rm ion}^0 \else $\xi_{\rm ion}^0$\fi}

\title[New insight on nebular \heii\ emission] 
{New insight on the far-UV SED and \heii\ emission from low metallicity galaxies
}

\author[D. Schaerer et al.]   
{Daniel Schaerer$^{1,2}$,
 Yuri Izotov$^3$,
  \and Tassos Fragos$^1$}

\affiliation{$^1$Observatoire de Gen\`eve, Universit\'e de Gen\`eve, 51 Ch. des Maillettes, 1290 Versoix, Switzerland \\ email: {\tt daniel.schaerer@unige.ch} \\[\affilskip]
$^2$CNRS, IRAP, 14 Avenue E. Belin, 31400 Toulouse, France\\[\affilskip]
$^3$Bogolyubov Institute for Theoretical Physics,
National Academy of Sciences of Ukraine, 14-b Metrolohichna str., Kyiv,
03143, Ukraine

}

\pubyear{2019}
\volume{xxx}  
\jname{Title of your IAU Symposium}
\editors{A.C. Editor, B.D. Editor \& C.E. Editor, eds.}
\begin{document}

\maketitle

\begin{abstract}
Understanding the ionizing spectrum of low-metallicity galaxies is of great importance for modeling and interpreting emission line observations of early/distant galaxies.

Although a wide suite of stellar evolution, atmosphere, population synthesis,
and photoionization models, taking many physical processes into account now exist,
all models face a common problem: the inability to explain the presence of nebular \heii\ emission, which is observed in many low metallicity galaxies, both in UV
and optical spectra. Several possible explanations have been proposed in the literature, including Wolf-Rayet (WR) stars, binaries, very massive stars, X-ray sources, or shocks. However, none has so far been able to explain the major observations.

We briefly discuss the \heii\ problem, available empirical data, and observed trends combining X-ray, optical and other studies.
We present a simple and consistent physical model showing that X-ray binaries could explain the long-standing nebular \heii\ problem.
Our model, described in \cite{schaerer2019}, successfully explains the observed trends and strength of nebular \heii\ emission in large samples of low metallicity galaxies 
and in individual galaxies, which have been studied in detail and with multi-wavelength observations. 
Our results have in particular important implications for the interpretation of galaxy spectra in the early Universe, which will be obtained 
with upcoming and future facilities.

\end{abstract}

\firstsection 
\section{The nebular \heii\ problem}
Nebular \Heii\ emission in optical spectra of nearby sources has been discovered in the late 1980ies (see e.g.\ \cite{Pakull1986Detection-of-an,Garnett1991He-II-emission-}).
Except for planetary nebulae (PN), where such emission is quite common, such objects are very rare in the local group, and the sources known then cover a diversity of objects, including a Wolf--Rayet star, a O type, a massive binary system, and one X-ray source (\cite{Garnett1991He-II-emission-}).
Since their discovery, the origin of this \heii\ emission has been puzzling. Only stars with very high effective temperatures $\teff \ga 80-100$ kK emit non-negligible amounts of \hep\ ionizing photons above 54 eV (e.g.\ \cite{schaerer2002}), and because such temperatures are only reached in very peculiar evolutionary 
phases (e.g.,\ in the WR or PN phase).

Zooming out of the local group, large numbers of star-forming galaxies showing nebular \heii\ emission have been found, and these high-ionization lines are also
present in AGN. E.g.\ using the Sloan survey, \cite{Shirazi2012Strongly-star-f} have found 2865 galaxies with nebular \heii. Excluding AGN,  they have  $\sim 200$
star-forming galaxies, some of them showing also the presence of broad emission lines, indicative of WR stars, in these galaxies. 
Since ``normal'' stars should be the most numerous ones in star-forming regions/galaxies,  and their UV/ionizing spectra are expected to be dominated
by massive stars, very hot stars (cf.\ above) are at best very rare in these objects. Therefore  no or very weak nebular \heii\ is expected in general.
In Figure 1 we show a compilation of more than 1400 star-forming galaxies or regions thereof showing nebular \heii\ emission from \cite{Izotov2016The-bursting-na}. This sample reprensents a significant jump in statistics. They are selected based on the quality of the spectra, allowing direct metallicity determinations
via the   auroral line method. Active galaxies are excluded based on  BPT diagrams.

Nebular \heii\ emission traced by the UV \Heiiuv\ line has also been observed in significant numbers of galaxies, some at low redshift and more at high-$z$ (e.g.\ \cite{Cassata2013He-II-emitters-}, \cite{Berg2018A-Window-on-the}, \cite{Nanayakkara}).
The same trend of increasing frequency and strength of nebular  \Heiiuv\ with decreasing metallicity seems to be observed (\cite{senchyna2017}),  as for the optical line. This is expected since the \heii\ lines are recombination lines from a simple hydrogenic atom, whose relative strengths between the UV and optical lines are to first order determined by atomic properties.

Since the discovery of \heii\ emitters, other sources and processes that could emit more energetic photons 
than stellar sources have been suggested. These include\ X-ray binaries (XRBs),  photoionization by X-rays, and strong shocks
(see e.g.\  \cite{Pakull1986Detection-of-an,Garnett1991He-II-emission-,Thuan2005High-Ionization,Kehrig2015The-Extended-He}).
Other studies have explored if rotation and binarity could significantly alter the evolution of massive stars and create sufficiently hot stars 
and hence a harder ionizing spectrum (e.g. \cite{szecsi}, \cite{Gotberg2018Spectral-models}).
%
None of them has so far been able to quantitatively explain the observed intensity of the \heii\ emission in low-metallicity galaxies. 
For example, the latest BPASS binary population and synthesis models underpredict the observed \Heii/\hb\ intensities
by $\sim$one order of magnitude, as shown in Fig.\ 2 (right).
Shocks and X-rays seem to be able to explain specific cases, but appear insufficient in others (see above references).
However, no predictive model that would allow linking shock models to other galaxy properties exists, 
and especially it is unclear how shocks would reproduce  the trends of \heii\ intensity with metallicity (Fig.\ 1).
In short, the \heii\ problem remains overall an unsolved problem calling for new/different approaches.

\begin{figure}[tb]
{\centering
\includegraphics[width=8cm]{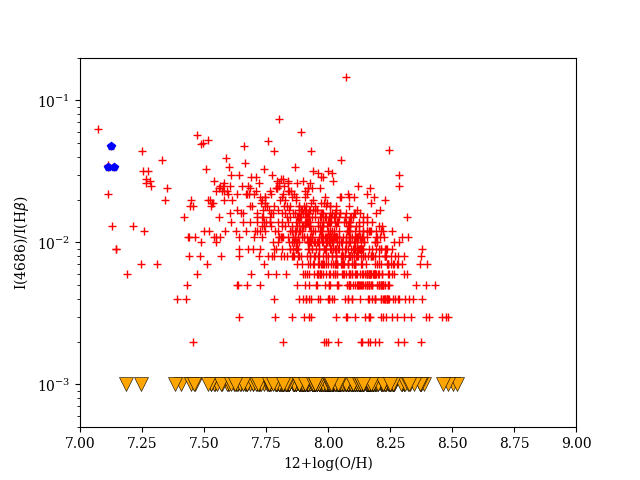}
\caption{Observed $I(4686)/I(\hb)$ relative nebular line intensities as a function of metallicity.
{\em Right:} Observations of low-metallicity star-forming galaxies from the SDSS DR12-14 and observations of Izotov and collaborators
are shown as  red plusses. Measurements of the NW region of I Zw 18 are plotted with blue diamonds.
The triangles show the non-detections ($\sim 1/3$ of the sources).
}}
\label{fig_heii_oh}
\end{figure}

\section{A fresh look at the \heii\ problem -- empirical results}
The literature on X-ray emission in and from star-forming galaxies (SFGs) provides interesting insights, which
could help understand the nature of \heii\ emission. Indeed, the observations show overall the following picture
(e.g.\ \cite{mineo2012}, \cite{Douna2015Metallicity-dep}, \cite{Brorby2016Enhanced-X-ray-}):
\begin{enumerate}
\item First, the X-ray emission of SFGs is dominated by point-like sources, whose luminosity follows a power-law, 
including ultra-luminous X-ray sources (ULX) at the bright end ($L_X > 10^{40}$ \ergs), 
high-mass X-ray binaries (HMXBs) and others. 
\item  The total (spatially integrated) X-ray luminosity correlates to {\em first order} with the total star formation rate (SFR).
\item There is an excess of X-ray emission per unit SFR, $L_X/$SFR, at low metallicity.
\item The observed scatter in $L_X/$SFR is due to stochasticity of the ULX+HMXB at the high luminosity end of the X-ray luminosity function, 
which dominates the integrated X-ray emission.
\end{enumerate}

These empirical findings suggest that X-ray bright sources (ULX and/or HMXB) could naturally explain the observed trends of 
\heii\ emission, in particular the increase of \heii/\hb\ with decreasing metallicity, for the following reasons and in the following way:
First, the \hb\  (or any other H recombination line) luminosity scales linearly with the SFR, i.e.\ $L(\hb)/SFR \approx$ const.
Second, assuming that bright X-ray sources emit a constant fraction of ionizing photons per X-ray luminosity above 54 eV, 
i.e.\ $q = Q({\rm He}^+)/L_X$, the number of ${\rm He}^+$ ionizing photons per $L_X$, is constant. Then the \heii\ line luminosity
is proportional to $L_X$, and the observed scaling of $L_X$ with SFR (point b) translates immediately to a relation between the  \heii/\hb\ 
line luminosities, hence also their relative intensity, which is the main observable. Finally, the observed increase of $L_X/$SFR towards low metallicity 
(point c) implies that $I(4686)/I(\hb)$ increases with decreasing O/H, as observed (Fig.\ 1). Furthermore, the observed scatter in $L_X/$SFR (point d)
would also lead to scatter in  $I(4686)/I(\hb)$. This is in essence what could explain nebular \heii\ at low metallicity, the observed metallicity trend, 
and possibly also the scatter, as we proposed in \cite{schaerer2019}. 

To relate the observed trends of nebular \heii\ with X-ray emission, we basically have to make only a single assumption, namely
that \heii\ emission originates from or is related to the bright sources which dominate in X-rays, i.e.\ ULX and/or HMBX. In fact,
this assumption appears very reasonable and is supported by some empirical evidence. Indeed \heii\ emission has been observed
in several ULX and bright HMXB nebulae (e.g. \cite{Kaaret2009}, \cite{Pakull2002}). Furthermore such sources are present in SFGs, 
and they significantly modifiy the ionizing spectrum (and hence affect certain emission lines), as shown e.g.\ in the very detailed study 
of the metal-poor \heii\ emitting galaxy I Zw 18 by \cite{Lebouteiller2017Neutral-gas-hea}. Indeed in this galaxy, Chandra observations have  
shown the presence of a ULX and demonstrated that it is located in the main \hii\ region (NW region), where the bulk of \heii\ emission
originates from. This strongly supports a physical connection between bright X-ray sources and nebular \heii. 
This finding, is also confirmed by a new study of \cite{Heap2019Radiative-signa}, who also show that a well-motivated ULX spectral 
model is able to quantitatively explain the \heii\ intensity in I Zw 18. 

In  \cite{schaerer2019} we have used the observed X-ray and \Heii\ line luminosities of I Zw 18 to empirically derive 
$q = Q({\rm He}^+)/L_X=(1.0-3.4)\times 10^{10}$ photon/erg and adopt a ``typical'' value of $q=2 \times 10^{10}$ photon/erg.
Predicted $I(4686)/I(\hb)$ relative intensities obtained using this simple approach are plotted in Fig.\ 2. 
They show that the observations can be reproduced approximately with these simple assumptions. 

\begin{figure}[tb]
{\centering
\includegraphics[width=6.7cm]{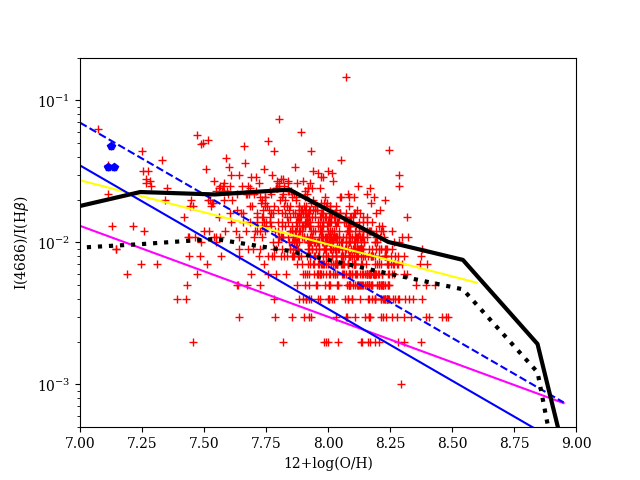}
\includegraphics[width=6.7cm]{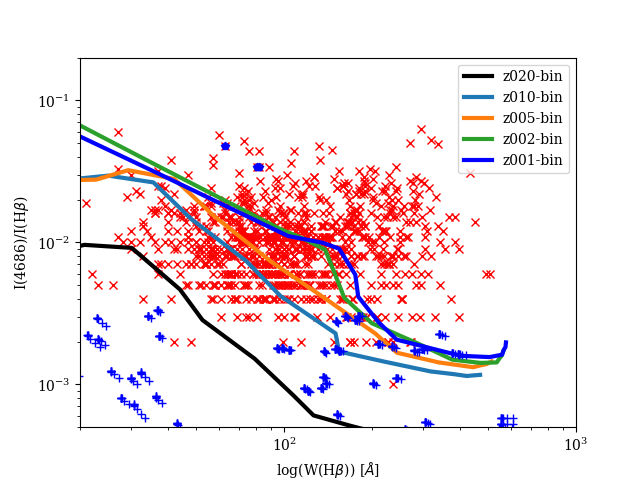}
\caption{Observed and predicted $I(4686)/I(\hb)$ relative nebular line intensities as a function of metallicity (left) and 
the \hb\ equivalent width (right) as an age indicator. Observations as in Fig.\ 1
{\em Left:} The observations are fitted by the yellow line (linear regression to the data points, detections only).
Assuming $q=2\times 10^{10}$ photon/erg, the empirical $L_X/{\rm SFR}$--O/H relations of 
\cite{Douna2015Metallicity-dep} and  \cite{Brorby2016Enhanced-X-ray-}  translate into \heii\ intensities 
shown by the blue and magenta solid lines, respectively, assuming constant SFR.
The black lines show the predicted \heii\ intensity adopting $L_X/{\rm SFR}$ predicted from 
the XRB synthesis models.
for a constant SFR over 10 Myr (dotted) and 0.1 Gyr (solid), and the same value of $q$.
The blue dashed line differs from the solid line by assuming a value of $q$ that is a factor of two higher.
{\em Right:} Instantaneous burst models for different metallicities, showing the predicted age dependence.
Blue crosses show the BPASS models from \cite{Xiao2018Emission-line-d}. Figs.\ from \cite{schaerer2019}.
}}
\label{fig_heii_oh}
\end{figure}

\section{Synthetic models including \heii\ emission from X-ray binaries}

To go beyond the simple empirical attempt, we have combined X-ray binaries (XRB) population synthesis models with models 
describing spectral evolution of  ``normal" stellar populations. The main results, taken from \cite{schaerer2019}, are shown in Fig.\ 2.

Concretely, we have used models developed by 
Fragos \etal\ (2013ab)
to study the cosmological evolution of XRB populations that were recently recalibrated to updated measurements of the cosmic star-formation history and metallicity evolution (\cite{Madau2017Radiation-Backg}.
An important prediction of these models is the strong dependence of the XRB population on metallicity, both
 in terms of the formation efficiency of XRBs and the integrated X-ray luminosity of the whole population.
 This means that at low O/H more XRBs are predicted and their X-ray luminosity is higher.
Therefore these XRB models predict a higher $L_X$/SFR at low O/H, as shown by the observations discussed 
above (point c). Additionnaly, these models  predict a strong dependence of $L_X$ on the age of the stellar population (see \cite{schaerer2019}).

To bracket a range of star formation histories, we have examined instantaneous bursts and constant SFR. The predicted  $I(4686)/I(\hb)$ intensity
are shown in Fig.\ 2. They broadly cover the range of the observations, showing that our models can fairly well reproduce the
observed  $I(4686)/I(\hb)$, possibly with a superposition of populations.

As already mentioned, binary populations synthesis models such as BPASS do not predict sufficiently hard ionizing spectra
to solve the \heii\ problem (cf.\ Fig.\ 2). This shows that even hot/rejuvenated stars, which are created through binary interactions,
are not sufficient, and that other sources (e.g.\ HMBX and ULX) need to be included.
The earlier synthesis models of \cite{cervino2002} including X-rays have also failed in this respect, since they primarily include soft X-ray emission from stellar winds, supernovae, and their remnants.

\section{Conclusions and outlook}

Based on empirical X-ray data for star-forming galaxies showing a very similar behaviour of increasing $L_X/SFR$ as nebular \heii\ emission
with metallicity (Fig.\ 1), on the observational finding of a ULX in the strong \Heii\ emitting galaxy I Zw 18, and on the
detailed modeling of this galaxy by \cite{Lebouteiller2017Neutral-gas-hea} and \cite{Heap2019Radiative-signa},
we have proposed that ULXs/HMXBs are the prime source of nebular \heii\ emission low metallicity SFGs.
A simple quantitative model using both empirical data and recent X-ray binary population synthesis models has recently been 
proposed by \cite{schaerer2019} to solve the long-standing problem of nebular \heii\ emission.

Obviously further applications, and tests are welcome and required, and this model does not exclude the 
contribution from other processes to the observed \heii\ emission.
For example, establishing observationally more firmly 
the correlation between X-ray
and \heii\ emission, both on spatially resolved scales and in integrated populations, should be  a useful test.
It also needs to be worked out if/how much the high-energy emission affects other emission lines, also in relation with
the apparent hardening of the SEDs at high-$z$ inferred by several studies  (see e.g.\ \cite{Stark2016Galaxies-in-the}).
In any case observations show that X-rays  are ubiquitous in star-forming galaxies (even if weaker than
in AGN), as discussed amply in the literature (cf.\ Sect. 1). Therefore, their effects on the SEDs need to be examined, 
a task which should be more important for high-$z$ and low-metallicity galaxies, since HMXBs, ULXs, and their X-ray
emission becomes stronger at low O/H.


\end{document}